# Characterization of a flat superpolished mandrel prototype with hard (TiN / SiC) overcoating to enhance the surface durability

D. Spiga (1,2)

*(1) Osservatorio Astronomico di Brera/INAF - Via Bianchi 46 – 23807 Merate (Lc) – Italy*
*(2) Università di Milano/Bicocca – Dipartimento di Fisica G.Occhialini – Piazza della Scienza 3 - 20126 Milano - Italy*

**ABSTRACT**

A number of hard X-Ray (10 - 100 KeV) astronomical missions of near future will make use of multilayer-coated focusing mirrors. The technology based on Nickel electroformed replication is suitable for the multilayer optics realization, since multi-modular telescopes are foreseen. For example, for the Constellation-X mission there is the need of realizing up to 14 identical modules (12 flight modules plus two spares) which can be replicated by the same series of mandrels. The Ni replication approach is derived from the method already successfully used for making the Au coated soft X-ray mirrors with good imaging performances of the missions BeppoSAX, XMM-Newton and Swift. In the technological extension of the process to the multilayer optics fabrication, it would be convenient to overcoat the external surface of mandrels (normally in Kanigen) with a layer made of a very hard material. This would help to maintain the very low roughness level requested by the application (typically less than a couple of Angstroms for a 1 micrometer scan length with AFM) also after many replications and successive cleaning of the mandrel. Good material candidate are at this regard TiN and SiC, both characterized by a very high hardness. We have proven that flat prototypes with TiN and SiC overcoating can be superpolished at a level comparable to the traditional electroless Nickel coating. In this paper we will present a characterization by topographic measurement (AFM and WYKO) and by X-Ray scattering of two of these samples.

## 1. INTRODUCTION

The extension to the hard X-ray energy window (10-100 KeV) foreseen by next focusing X-ray telescopes will require the use of grazing incidence mirrors with very shallow angles (0.2 – 0.05 deg). The mirror shells will be coated by multilayer reflecting films or with a high-density single-layer (e.g. Iridium). In any case, due to the modest aperture (35 cm at most with 10 m focal length) diameters imposed by the small reflection angles with focal length up to 10 m, the hard x-ray telescopes have to be based on multimodular optics with many nested mirror shells, in order to get an acceptable collecting area. This implies that a number of identical mirror shells have to be produced.

In recent years a lot of improvements have been reached in the field of vacuum deposition techniques to be applied to the case of X-ray multilayer mirrors and the feasibility of this kind of optics is nowadays considered mature. In this context, a suitable technique for the realization of the mirror shell is that based on the Nickel electroforming replication from shaped mandrels. This technique has already been successfully used to make the soft X-ray optics of satellites like Beppo-SAX[1], SWIFT[2], Newton-XMM[3], operating with single-layer Au coated mirrors. The use of the replication approach would mean a reduction of realization cost, time, and would make easier an industrialization process, while keeping the optical performances at a good level. In fact, in addition to a very high throughput, it is well known that Nickel electroformed mirrors are characterized by good imaging properties. Moreover this process well satisfies the requirements of multimodular optics since, the same master mandrel can be used for making all shells of a given diameter for all modules. Finally, the use of monolithic shells greatly simplify the assembly procedure with respect other production methods under development based e.g. on segmented substrates. Replication by Ni electroforming is

---

· e-mail: spiga@merate.mi.astro.it

not only the technology assumed for making the optics of the balloon projects HERO[4] and HEXIT[5], but it is also a good candidate for the production of the hard X-ray mirrors for the hard X-ray experiment HXT aboard Constellation-X[6].

As already mentioned, such technique may be extended to replicate multilayer-coated mirrors. On the other hand, it is well known how the microroughness of the reflecting surface hampers the x-ray mirror reflectivity: this is an even more crucial point in multilayer–coated mirrors, because the reflected intensity falls down exponentially at every layer with its interface rms. The interface smoothness is then an essential point to care about. This goal can be reached by a careful study and optimization of the deposition process, but also by depositing the layers onto a mandrel whose surface has been superpolished at excellent level. The film quality is in fact very sensitive to initial defects of the surfaces where they are deposited on, and as long as layers grow, the interface roughness could be amplified up to unacceptable values[16, 17].

It is then a fundamental step to produce mandrels having a very low surface microroughness. Our requirement to be met is an rms $\sigma < 2.5$ Å in the wavelength range $0.1 - 10$ μm. Such levels are going to be reached with developments performed in our labs by adopting usual Nickel coated mandrels. Another respect has, however, to be taken into account. The replication of a series of shells by using a mandrel tends to deteriorate its surface and after a few replication processes the quality of the mandrel surface could be compromised. Moreover, at every replication residual particles of the film usually remain in place onto the mandrel surface and, after a few iterations, must be removed. During this cleaning phase the mandrel may be damaged, and its microroughness consistently raises. In order to solve these problems and in order to not to be forced to repeat the long and expensive superpolishing process, it is convenient to overcoat the mandrel with a very hard material, provided that it can be polished at similar levels as electrochemical Nickel. Due to their very high structural features (in particular the hardness), materials like the Titanium Nitride (TiN) and the Silicon Carbide (SiC) seem to have the requested properties. In this paper we will describe the characterization performed on flat prototype mandrels with TiN and SiC overcoatings, showing that they can be superpolished at microroughness levels comparable to the usual Ni mandrels. These materials are therefore optimal candidate to be used in the realization of master mandrels for future hard X-ray optics.

## 2. REPLICATION TECHNIQUE

We are at present time concentrating on two different ways to produce hard X-ray optics, both derived from the usual Nickel electroforming technique[9] (fig. 1). The first step is, in both cases, the production of an Aluminium-made mandrel which represents the negative profile of the shell of the x-ray mirror. The mandrel is then electrochemically coated by a 100 μm layer of Nickel (Kanigen), a material suitable to be highly polished due to its amorphous structure.

Using the first method (Fig. 1-a), the multilayer film is directly deposited under vacuum on the mandrel and then the mirror wall (Ni) is electroformed on the surface of the last deposited layer. The mirror (multilayer plus Ni substrate) separation from the mandrel is obtained by cooling, thanks to the large difference (about a factor two) in coefficients of thermal expansion of Aluminium and Nickel. The second way (see Fig. 1-b) is based on the replication from the mandrel of a Au-coated Nickel carrier, followed by the deposition of the multilayer film by a linear deposition source onto the carrier internal surface. Development carried out for both methods have dealt to very encouraging results[10].

Table 1. Roughness levels (rms) at different scan lengths as measured for the mandrel #12 of the Beppo-SAX series, compared to a Ni coated mandrel, superpolished with the new procedure[11].

| Instrument used for the surface scanning | Scan Length (μm) | Roughness (rms) for mandrel SAX #12 (Å) | Roughness (rms) for a Ni superpolished mandrel (Å) |
|---|---|---|---|
| *WYKO – 20 X* | 660 | 7.6 | 3.0 |
| *AFM* | 10 | 6.2 | 2.4 |
| *AFM* | 1 | 3.4 | 1.8 |

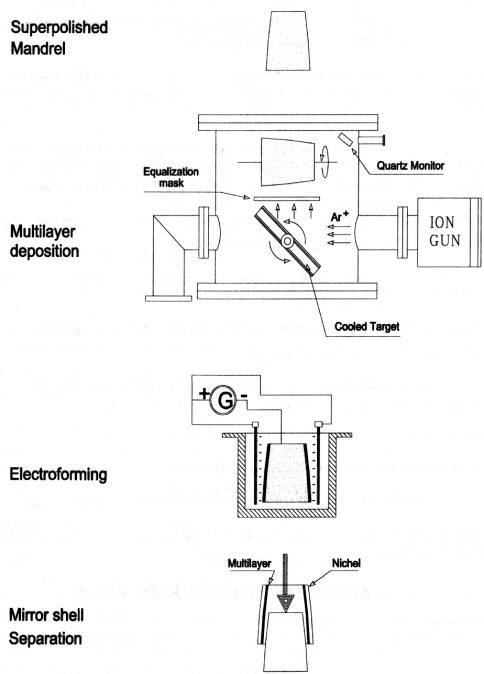
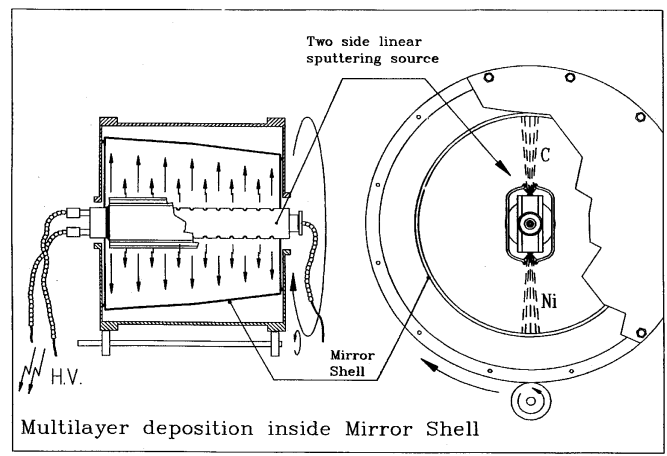

**(A)**          **(B)**

**Fig.1 X-Ray mirror production by replication : (A) direct deposition of the multilayer film onto the surface of a superpolished mandrel and successive replication by Ni electroforming; (B) application of the multilayer film onto the surface of a previously Ni replicated Au-coated substrate by using a vacuum deposition system based on a linear source.**

As already discussed, it is essential to start from a superpolished mandrel. Therefore, the mandrel has to be polished at a roughness level much better than in the case of simple Au-coated optics used for soft X-ray optics. This is due to the amplification of roughness at each layer in multilayer growth. At OAB-INAF we developed a new lapping process in the aim of enabling superpolishing of mandrels at the requested level. In Tab. 1 the microroughness (rms values) for a mandrel taken from the BeppoSAX series and a prototype polished with the new technology are reported. The gain is evident.

## 3. CHARACTERIZATION OF FLAT PROTOTYPES WITH HARD OVERCOATING

We have characterized two flat prototypes made of Aluminium, at which a TiN and a SiC overcoating were respectively applied. In the first case (TiN) the sample was first overcoated with electroless Nickel, which was polished before the overcoating of the hard material at a level of ten angstrom rms as measured with a WYKO 20x optical profilometer, in order to avoid print-through effects in the final superpolished phase. This protoype is a 10 cm diameter disk, coated by a few microns TiN layer by reactive sputtering, and then superpolished (at OAB) again down to few angstrom of rms.
The SiC prototype is instead a 2.5 cm diameter Aluminium disk, coated with a 18 μm SiC layer at Ce.Te.V. by PECVD method and then superpolished at ZEISS (Germany) by using diamond powders, needed due to the extreme hardness of the material.
One of the mainly relevant properties of TiN and SiC are their extreme hardness and stability (see tab. 2) and for these reasons they are optimal candidates for the overcoating of mandrels to be used in x-ray mirror replication. Moreover, TiN (a conductor) and SiC (an insulator) are excellent non-sticky materials, which makes easier the mirror separation at the last step of the replication procedure. They are also inert to acids, bases, solvents, caustic.

Table 2: comparison between properties of suitable materials for mandrel coatings
(data taken from www.brycoat.com and www.goodfellow.com)

| Parameter | Nickel | Titanium Nitride | Silicon Carbide |
|---|---|---|---|
| Vicker Hardness (Kgf/mm$^2$) | 721 | 1300 - 2000 | 2500 – 3100 |
| Density (g/cm$^3$) | 8.9 | 5.22 | 3.1 |
| Young's modulus (GPa) | 214 | 600 | 200 – 400 (ceramic) |
| Melting point ($^o$C) | 1455 | 2930 | 2650 – 2950 |

For the characterization of our samples we used topographic instruments and X-ray scattering. The topographic measurements were performed by using a stand-alone Atomic Force Microscope DIII by Digital Instruments able to provide 3-D maps with scan length of 100, 10, 1 μm, and a WYKO optical profilometer TOPO 2D in the 20x magnification which provides monodimensional 600 μm-wide scans with 1024 sampled points. Finally, in these study, we even made use of Nomarski Contrast Phase Microscopy, which allowed us to acquire a direct optical images of the surfaces. Concerning the x-ray scattering measurements we used a BEDE X-Ray Diffractometer[21] at the wavelength of 1.541 Å (Cu Kα1 line) and 0.709 Å (Mo Kα1 line).

### 3.1 Analysis of the scattering data

The scattering data analysis is performed by analyzing the distribution of an x-ray beam reflected by the sample. The photons are mainly reflected in the specular direction but some are also scattered in its vicinity by the surface microroughness.

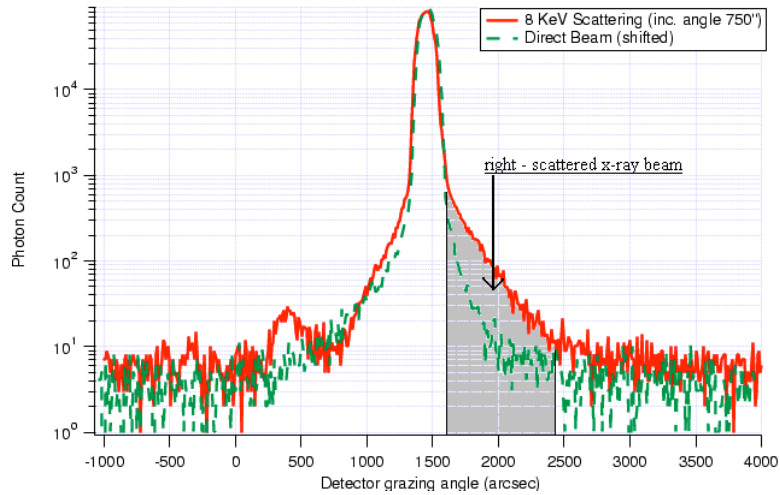

Fig. 2. A 8.05 KeV beam scattered by the TiN sample surface at 750'' of grazing incidence angle.

The scattering of the x-ray beam in vicinity of the specular reflected beam can be used to compute the surface Power Spectral Density (PSD): the perturbation theory[18, 19, 20] in smooth-surface approximation yields this relationship between the Scattered-Intensity angular distribution I(θ) and the surface PSD P(θ) which scatters the λ wavelength x-rays at the grazing angle $θ_s$:

$$\frac{1}{I_0}\left(\frac{dI}{dθ_s}\right) = 4 k^3 R(θ_i, θ_s) \operatorname{sen}θ_i \operatorname{sen}^2 θ_s P(θ_s)$$

where $k = 2\pi/\lambda$, $I_0$ is the incident beam on the sample and $\theta_i$ the grazing incidence angle. Usually only the right-scattered beam is considered and a further factor 2 appears at the left-hand term of the equation. The term $R(\theta_i, \theta_s)$ is a function of the scattering and the incidence angle. This function is derived from the assumed theory (scalar, vector, smooth-surface, rough-surface...), but in our wide scattering angle sensitivity range and as the smooth surface requirement is met by all of our samples, we can restrict to the Rayleigh-Rice vector theory, which yields:

$$R(\theta_i, \theta_s) = [R(\theta_s) / R(\theta_i)]^{1/2}$$

being $R(\theta_s)$ and $R(\theta_i)$ the reflectivities evaluated at the scattering and incidence angles, respectively. Furthermore, the scattering angle can be related to an associated spatial wavelength $l$ of the surface roughness, by the first-order grating equation:

$$l = \frac{\lambda}{|\cos\theta_s - \cos\theta_i|}$$

The specular direction would correspond to the zero frequency, that is an ideal smooth surface. In practice, because of the finite x-ray spatial instrument sensitivity, it is impossible to reach this limit, and we will have a lower limit in the explorable frequency range. Similarly, in the wide-angle scattering region, as the scattered intensity falls down the instrument average noise level, there we can locate our wavelength band upper limit.

The scattering profile allows us to relate the photon distribution to the surface PSD as a function of spatial wavelength. That is, to give a complete characterization of the surface in the limits imposed by the instrument spatial resolution and by the instrumental noise. which should be in agreement to the derived PSD from WYKO anf AFM topographic measurements. The squared rms (referred to the sensitivity frequencies window) is simply obtained by PSD spatial frequencies integration.

## 4. TITANIUM NITRIDE MICROROUGHNESS MEASUREMENTS

The TiN sample at the Nomarski Microscope showed some point-like defects in ejection with typical diameter of a few microns and a typical height of 20 nm (they are also notable in $(100~\mu m)^2$ scan of AFM), which cover a fraction of the surface depending on the considered point (1% at least, but in some points they are much more numerous): these defects and their polishing are still under investigation. There are also some scratchs due to the powder-lapping process, but not affecting in a sensitive way the microroughness level of the surface.

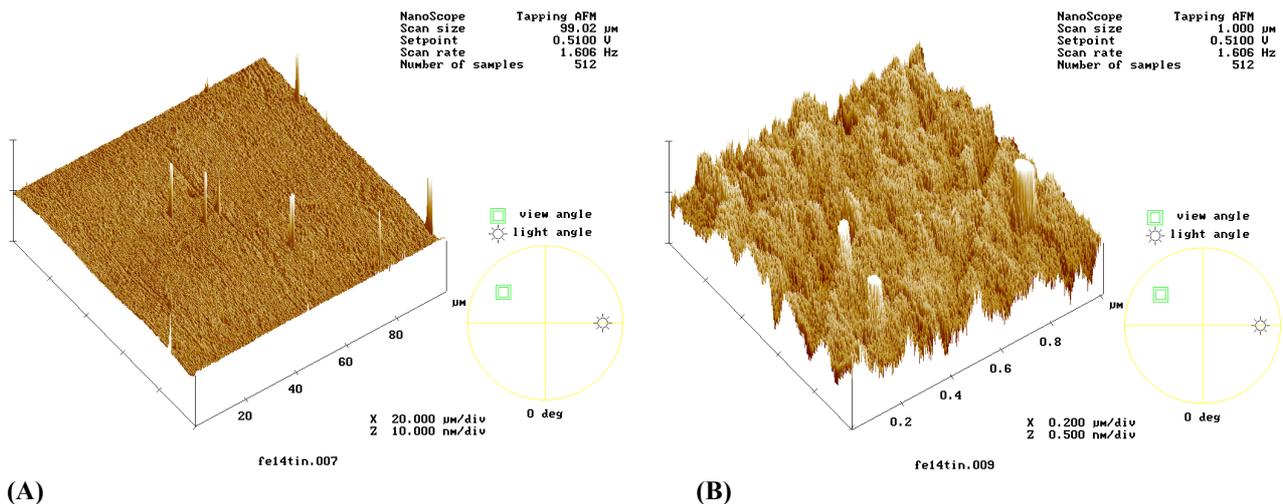

(A)          (B)

**Fig. 3. AFM digital maps images of the TiN superpolished sample. (A) 100 μm scan length, the scale for the ordinate axis is 10 nm/div. (B) 1 μm scan length, the scale for the ordinate axis is 0.5 nm/div.**

In order to get a quantitative topographic surface analysis, AFM scans have been taken in various surface points. The defects visible in the Nomarski image are evident also in the 100 μm scan (fig. 3-a : σ = 2.7 Å), while the background surface seems to be quite smooth. A 10 μm and 1 μm scan (fig. 5-b: σ = 0.8 Å) show the nature of the low-amplitude roughness. These roughness value show a good improvement in smoothness with respect to the previously reached smoothness with Beppo-SAX (see Tab.1). The same improvement is observed also in the WYKO 20x profiles (σ = 3.3 Å). Moreover, by computing the 1-D PSD of these scans (which approximate power laws, as predicted by the theory[20] for fractal surfaces), we can also conclude that the measurements are in agreement each other (see fig.4) as the PSD are superposed.

As explained in the previous section, another valuable method we have used is the x-ray scattering analysis. The scattering measurements allow us to measure the surface PSD without informations about the chemical composition of the sample. The performed scattering measurements are shown in fig. 4: the measured PSD by XRS (either at 8,05 or 17.4 KeV) are in a good agreement to the WYKO and AFM data .

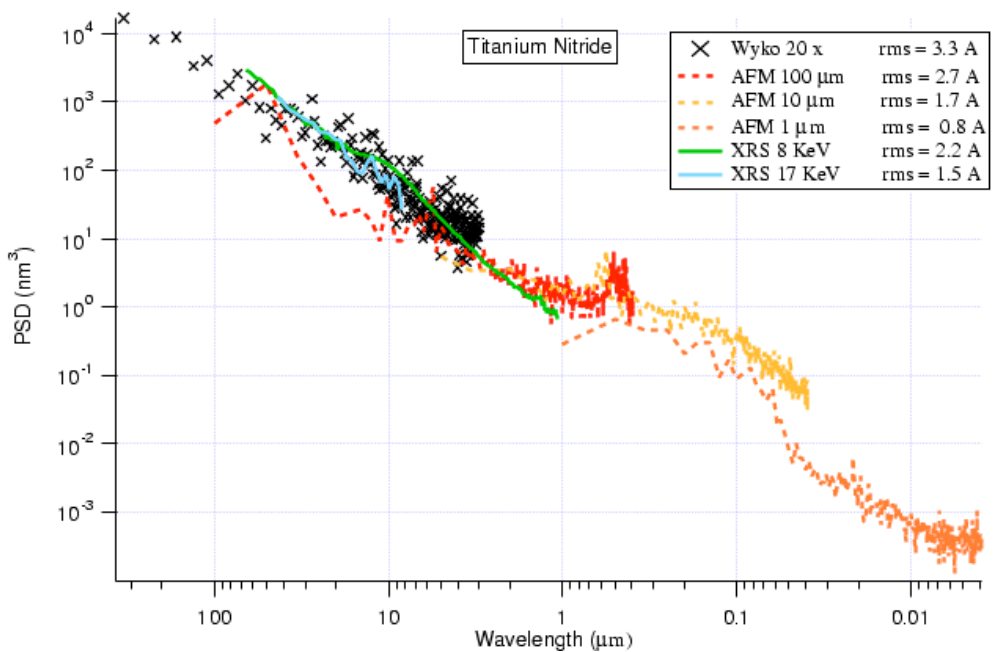

**Fig. 4. PSD results from 8 and 17 KeV X-ray scattering (solid lines), from AFM scans (dashed lines) and from WYKO profilometer (black crosses).**

As already mentioned, the adoption of hard overcoatings is suggested by the need of reusing the same mandrel to replicate a mirror shell many times as the requested number of x-ray mirror modules. At the end of some replication the residual particles of the deposited film must be removed in order to avoid the reflectivity degradation in following replications. In order to understand whether our coating can effectively stand the deposition followed by a removal of the deposit, the TiN sample has been coated with a 300 Å-thick gold layer by e-beam deposition (at Media Lario, Bosisio Parini, Italy). This Au layer has then been removed from the sample by tape lift. The residuals of gold and tape have then been ultimately removed with an Acetone cleaning. After these steps, we have measured again the sample by X-ray reflectivity and scattering.

The reflectivity results are shown in figure 5: the reflectivities before and after the Au coating and removal are identical, up to thousands of arcsec grazing incidence angles. As the reflectivity of a superpolished surface is strongly determinated by the its roughness state, we can conclude that the smoothness of the sample was not damaged by a deposition and separation of a layer which simulate, in fact, the operative conditions of mandrel replication. Future measurements will help us to understand whether repeated Au removals do not change the TiN surface smoothness.

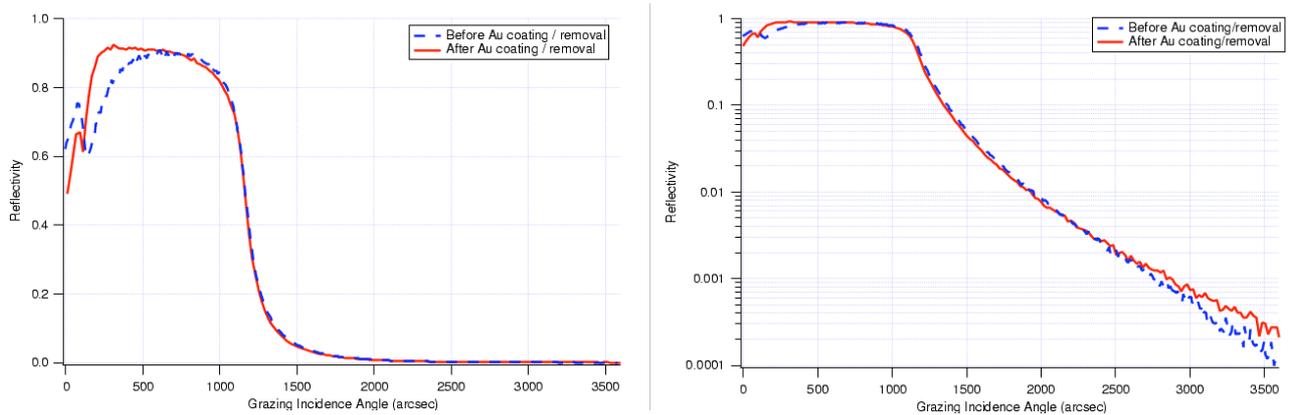

**(A)** **(B)**
**Fig. 5. TiN 8.05 KeV reflectivity profiles before (dashed line) and after (solid line) the deposition and removal (by tape lift) of the Au layer. Linear plot (A) and logarithmic plot (B)**

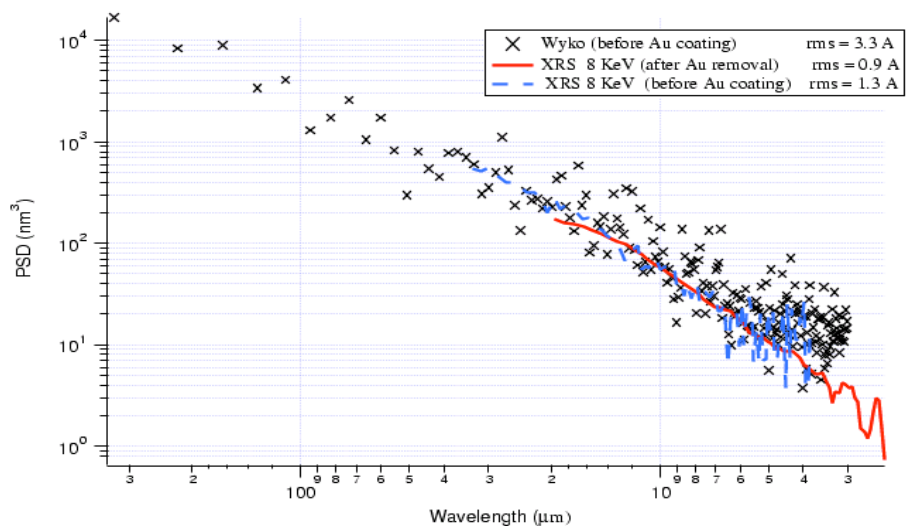

**Fig. 6. PSD comparison in TiN before (dashed line) and after (solid line) the Au coating deposition and removal. The black crosses represent the WYKO measurements before the deposition.**

A further confirmation of the resistance of TiN overcoating to the separation of a deposited layer comes from the PSD comparison. The x-ray scattering at 8.05 KeV has been measured on the surface after Au removal and compared to the surface PSD before the deposition. The result is plotted in the fig. 6. The two PSDs are almost superposed, and at no measured wavelength the PSD of the surface after deposition has a larger value than the PSD previously measured. This is another evidence that the surface characteristics were unchanged following the deposition/removal process.

## 5. SILICON CARBIDE MICROROUGHNESS MEASUREMENTS

The SiC sample images shows a less number of point-like defects than TiN. Only very weak scratches due to the diamond powder used for the superpolishing are visible in the Nomarski photos. On the other hand, AFM 10 µm scans show (see fig. 7) some undulations in the surface of 2-3 µm wavelength, superposed to some low point-like defects. However, the height of these features is less than 2 nm. The roughness rms is 3.7 Å with a 10 µm scan and 0.8 Å with a 1 µm scan.. The PSD measured by the AFM scans are confirmed by their good superposition.

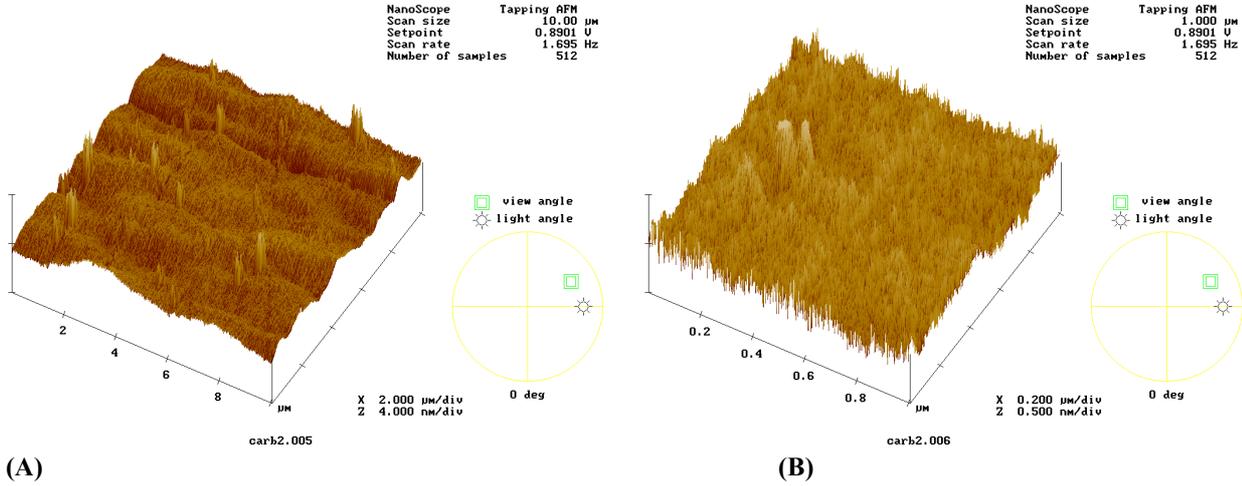

**(A)**  **(B)**
**Fig. 7.** AFM digital maps images of the SiC superpolished sample. (A) 10 µm scan length, the scale for the ordinate axis is 4 nm/div. (B) 1 µm scan length, the scale for the ordinate axis is 0.5 nm/div.

Because of the limited sample size, the performed x-ray measurements allow us to explore only the largest spatial wavelength window (the lower limit is 20 µm). Within this limit, the x-ray scattering measurement superposes quite well to the 100 µm AFM PSD (see fig. 8). Future planned measurements ought to allow us the extension of the XRS measurements at larger angles, which could give us a confirmation of the topographic measurements at lower spatial wavelength.

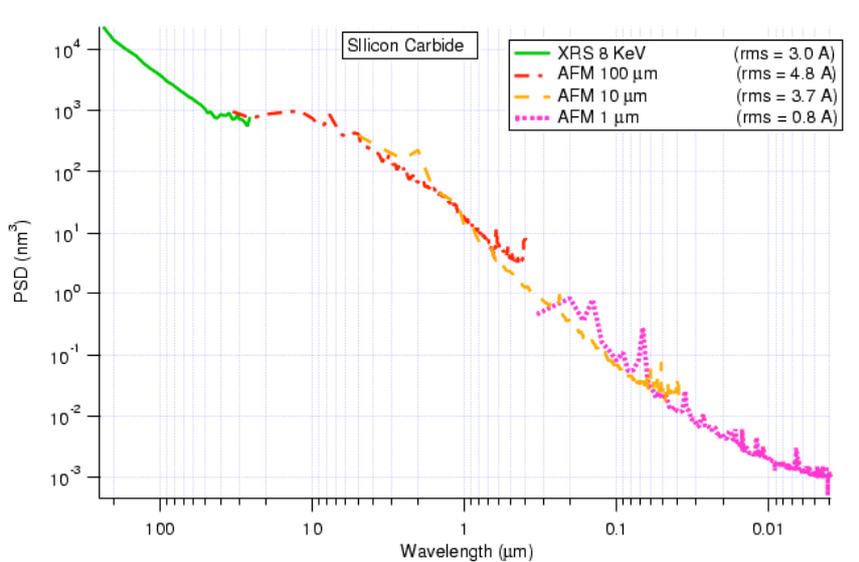

**Fig. 8.** SiC PSD results from 8 KeV X-ray scattering (solid lines) and from AFM scans (dashed lines).

## 6. CONCLUSIONS AND FINAL REMARKS

We did the characterization of two flat superpolished samples with hard overcoating material in TiN and SiC to determine the achieved smoothness of its surface. These tests allow us to understand the effectiveness of the polishing procedure onto hard materials. The microroughness measurements performed on the flat samples showed that a multi-instrument analysis lead us to a coherent result. The TiN and SiC samples can be polished at comparable levels as the already experimented Nickel (the usual external material of replication mandrels), and thanks to their hardness properties they are suitable materials for the mandrel replication. The x-ray scattering test has been done also on a Nickel flat sample superpolished following the same superpolishing procedure adopted for TiN. The resulting PSD is plotted in figure 9 in comparison to some PSD for TiN and SiC. Although the roughness rms might seem less for Ni than for TiN, the PSD are similar. The better smoothness of Ni is mainly due to the less extent of spatial frequency of the measurement. Moreover, we can observe that the PSD values of TiN are less than Ni in the region around 1 μm, which is the wavelength range that mainly amplifies the surface defect growth in multilayer films. We can conclude that the superpolishing procedure works well for TiN as for Ni (at least in the explored frequency range), which was since now the ideal surface coater that could be polished at the requested levels by x-ray optics.

The SiC PSD is not still at the same level as the Ni. The PSD comparison shows that in the 1-5 μm wavelengths range the SiC has a higher roughness than Ni and TiN. The situation changes, however, around some tenth of micron, where the SiC PSD falls down the TiN. This means that it is possible to attain superpolishing levels in SiC which are comparable to TiN at least in the 0.1- 0.5 μm wavelengths. In the SiC case an improvement of the superpolishing method is then necessary, in order to lower the PSD level in the 1 μm region as we did with TiN.

**Table 3. Comparison among the obtained microroughness of our flat samples. The XRS measurement rms are not directly comparable because each of then is referred to a different frequency range.**

| Used Instrument | Nickel (Å) | Titanium Nitride (Å) | Silicon Carbide (Å) |
|---|---|---|---|
| WYKO | 3.0 | 3.3 | N.A. |
| AFM 100 μm | N.A. | 2.7 | 4.8 |
| AFM 10 μm | 2.4 | 1.7 | 3.7 |
| AFM 1 μm | 1.8 | 0.8 | 0.8 |

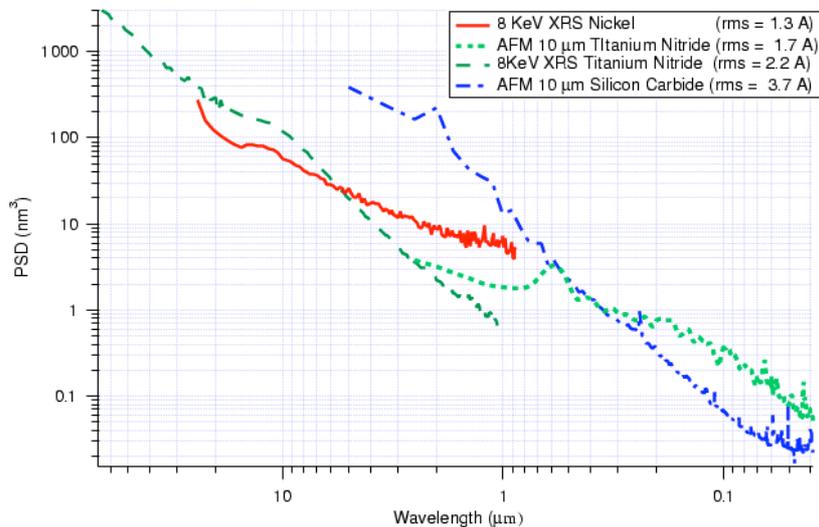

**Fig. 9.** Some PSD results with superpolished TiN sample (dashed line) and SiC sample (dashed-dot line) compared to a Ni sample (solid line).

Further measurements on these materials will provide an extension of the characterization. A positive result could lead to the production of hard-overcoated mandrels.

## ACKNOWLEDGEMENTS

Many thanks to the guide and collaboration of G. Pareschi and O.Citterio. The valuable collaboration by S. Cantù, S. Basso, and R. Valtolina (OAB-INAF) is acknowledged. We thank Ce.Te.V. (Carsoli, Italy), and in particular A. Mengali, for the PECVD deposition of the SiC sample. We are grateful to Media Lario (Bosisio Parini, Italy), and in particular to G. Valsecchi, for the e-beam deposition of Au onto the TiN sample. The TiN and Ni superpolishing process has been performed by P. Cerutti (Porto Valtravaglia, Italy). The SiC superpolishing has been made by ZEISS. This research is funded by the Italian Space Agency (ASI).

## REFERENCES

1. O. Citterio, G. Bonelli, G. Conti, E. Mattaini, E. Santambrogio, B. Sacco, E. Lanzara, H. Brauninger, and W. Burkert, "Optics for the X-ray imaging concentrators aboard the X-ray astronomy satellite SAX", *Appl. Opt.*, **27**, 1470, (1988).
2. D. N. Burrows, J. E. Hill, J. A., Nousek, A. A. Wells, A. D. Short, R. Willingale, O. Citterio, G. Chincarini, and G.Tagliaferri, "Swift X-Ray Telescope", *SPIE Proc.*, **4140**, 64 (2000).
3. F. A. Jansen, "XMM: advancing science with the high-throughput X-ray spectroscopic mission", ESA Bulletin, **100**, 15 (1999).
4. B.D. Ramsey et al. "HERO: high-energy replicated optics for hard x-ray balloon payload", *SPIE Proc.*, **4138**, 147 (2000)
5. G. Pareschi et al. "The HEXIT (High Energy X-ray Imagng Telescope) balloon- borne mission", Proc. of the 16th ESA Symposium on European Rocket and Balloon Programmes and Related Research", St. Gallen (Ch) 2-5 June 2003, in press
6. F. Harrison et al., "Technology development for the Constellation-X Hard X-Ray Telescope", *SPIE Proc.*, **3765**, 104 (1999)
7. P. Gorenstein, A. Ivan, R.J. Bruni, S. E. Romaine, F. Mazzoleni, G. Pareschi, M. Ghigo, and O. Citterio, "Integral shell substrates for the Con-X hard x-ray telescope system", *SPIE Proc.*, **4138**, 10 (2000).
8. O. Citterio, P. Cerutti, F. Mazzoleni, G. Pareschi, E. Poretti, P. Laganà, A. Mengali, C. Misiano, F. Pozzilli, and E. Simonetti, "Multilayer optics for hard X-ray astronomy by means of replication techniques", *SPIE Proc.*, **3766**, 310 (1999).
9. G. Pareschi, O. Citterio, M. Ghigo, F. Mazzoleni, "Replication by Ni electroforming approach to produce the Con-X/HXT hard X-ray mirrors", *SPIE Proc.*, **4851**, 528 (2002).
10. S. Romaine et al. "Development of a prototype Nickel optic for the Con-X hard x-ray telescope", *SPIE Proc*, **5168**, in press (2003, present issue)
11. Citterio et al., "Development of of soft and hard X-ray optics for astronomy: progress report 2 and considerations on material properties for large-diameter segmented optics of future missions", *SPIE Proc.*, **4496**, 23 (2001)
12. *Science with the Constellation-X Observatory*, NASA Publ. **NP-19998—067-GSFC** (1999) – See also the Constellation-X WWW page at: http://constellation.gsfc.nasa.gov/.
13. O. Citterio, M. Ghigo, F. Mazzoleni, G. Pareschi, L. Peverini, "Development of soft and hard X-ray optics for astronomy", *SPIE Proc.*, **4138**, 43 (2000).
14. S. E. Romaine, A. M. Hussein, J. Everett, A. M. Clark, R. J. Bruni, P. Gorenstein, M. Ghigo, F. Mazzoleni, and O. Citterio, "Application of multilayer coatings to replicated substrates", *SPIE Proc.*, **3444**, 564 (1998).
15. "Nitruri di Titanio" CeTeV (Centro Tecnologie del Vuoto) report
16. D. G. Stearns, D.P. Gaines, D.W. Sweeney and E.M. Gullikson, "Nonspecular x-ray scattering in a multilayer-coated imaging system", Journal of Applied Physics, **84**, 2 (1998)
17. D. G. Stearns et al., "Defects from substrate particles depend on the sputter deposition process", Solid State Technology, nov 2000, 95
18. E. L. Church, "Role of surface topography in x-ray scattering", *SPIE* **184** Space Optics, 196-202 (1979)

19. E. L. Church and P. Z. Takacs, "The interpretation of glancing incidence scattering measurements", *SPIE Proc.*, **640**, 126-133 (1986)
20. J.C. Stover, "Optical Scattering: measurement and analysis", *SPIE Press* (1995).
21. N. Locksley, B.K. Tanner, D.K. Bowen, "A novel beam conditioning monochromator for high-resolution x-ray diffraction", J. Appl. Cryst., 28, 314 (1995)